\documentclass[fleqn,10pt]{wlscirep}
\usepackage[utf8]{inputenc}
\usepackage[T1]{fontenc}
\usepackage{ascmac}
\usepackage{subcaption}
\usepackage{here}
\title{Assessment of image generation by quantum annealer}
\author[1*]{Takehito Sato}
\author[1,2,3]{Masayuki Ohzeki}
\author[1]{Kazuyuki Tanaka}

\affil[1]{Graduate School of Information Sciences, Tohoku University, Sendai, Japan}
\affil[2]{Institute for Innovative Research, Tokyo Institute of Technology, Yokohama, Japan}
\affil[3]{Sigma-i Co., Ltd., Tokyo, Japan}

\affil[*]{takehito.sato.t3@dc.tohoku.ac.jp}

\begin{abstract}

Quantum annealing was originally proposed as an approach for solving combinatorial optimisation problems using quantum effects.
D-Wave Systems has released a production model of quantum annealing hardware. 
However, the inherent noise and various environmental factors in the hardware hamper the determination of optimal solutions.
In addition, the freezing effect in regions with weak quantum fluctuations generates outputs approximately following a Gibbs--Boltzmann distribution at an extremely low temperature.
Thus, a quantum annealer may also serve as a fast sampler for the Ising spin-glass problem, and several studies have investigated Boltzmann machine learning using a quantum annealer. 
Previous developments have focused on comparing the performance in the standard distance of the resulting distributions between conventional methods in classical computers and sampling by a quantum annealer.
In this study, we focused on the performance of a quantum annealer as a generative model. 
To evaluate its performance, we prepared a discriminator given by a neural network trained on an a priori dataset.
The evaluation results show a higher performance of quantum annealing compared with the classical approach for Boltzmann machine learning.

\end{abstract}
\begin{document}

\flushbottom
\maketitle
%
%
\thispagestyle{empty}
\section*{Introduction}
Quantum annealing (QA) is an approach for solving combinatorial optimisation problems using quantum effects \cite{Kadowaki1998}.
The hardware implementation of QA has been developed by D-Wave Systems and performs for production level, attracting much interest from academia and industry \cite{Dwave2010a,Dwave2010b,Dwave2010c,Dwave2014}.
Quantum annealers have been used in numerous applications, such as portfolio optimisation \cite{Rosenberg2016}, protein folding \cite{Perdomo2012}, molecular similarity problems \cite{Hernandez2017}, computational biology \cite{Richard2018}, job-shop scheduling \cite{Venturelli2015}, traffic optimisation \cite{Neukart2017}, election forecasting \cite{Henderson2018}, machine learning \cite{Crawford2016,Arai2018nn,Takahashi2018,Ohzeki2018NOLTA,Neukart2018,Khoshaman2018}, web recommendation \cite{Nishimura2019}, and automated guided vehicles in factories \cite{Ohzeki2019}.
By feeding a Hamiltonian-based quadratic unconstrained binary optimisation matrix into a quantum annealer, several solutions can be quickly obtained. 
However, the solution may not always be optimal owing to unavoidable hardware limitations.
For instance, connectivity in the hardware graph may be insufficient to represent the optimisation matrix.
In fact, the original optimisation problem is embedded in a hardware chimaera (previous hardware version) or a Pegasus graph.
The embedding protocol is generally difficult, and many variants have been proposed \cite{Okada2019}.
When various problems are embedded in the hardware graph, a chain with interacting redundant spins represent the original problems.
Nevertheless, when the strength of the chain interaction is sufficiently high, embedding can be avoided.
Another hardware limitation is the nonuniform distribution of the degenerate ground states owing to quantum effects.
As an alternative for QA, simulated annealing achieves uniform degenerate ground states \cite{Matsuda2009,Yamamoto2020}.
Still, this limitation is not critical, as a degenerate ground state can be found in any case.
Another limitation is relatively crucial and related to the freezing effect and environment effect from a heat bath.
A real quantum annealer is not isolated from the environment. Thus, it does not reflect the ideal QA case assumed in theory.
Hence, if the system is thermalised, the QA outputs follow a Gibbs--Boltzmann distribution.
Consequently, several protocols based on QA do not maintain the system in the ground state as in adiabatic quantum computation. 
Instead, they employ a nonadiabatic counterpart \cite{Ohzeki2010a,Ohzeki2011,Ohzeki2011proc,Somma2012} and consider thermal effects \cite{Kadowaki2019}. In addition, environmental effects cannot be avoided.

Instead of finding optimal solutions, the quantum annealer may be used to generate outputs following a Gibbs--Boltzmann distribution.  
In this case, another problem of quantum annealers occurs.
Experimentally, the energy of solutions attained by a quantum annealer follows a Gibbs--Boltzmann distribution with finite-strength quantum fluctuations \cite{Amin2015}.
Although many QA applications have been reported, most of them focus on optimisation, and only a few developments have addressed QA applications of Boltzmann machine learning using quantum annealers.
Previous studies have shown that the quantum annealer achieves higher performance in Boltzmann machine learning measured in the standard distance between the target and attained distributions, the Kullback--Leibler (KL) divergence.

In this study, we investigated the performance of Boltzmann machine learning using a quantum annealer from a different perspective.
We used the quantum annealer during training and generation of the Boltzmann machine.
Previously, the performance of the quantum annealer has been investigated with respect to the final value of the KL divergence.
However, the trained model has not been considered as a generative model, as its performance is difficult to assess.
Thus, we prepared another neural network to measure the quality of the generated data from the Boltzmann machine.
Although several measures quantify the similarity between two images, we use another neural network because the generated outputs do not necessarily represent a one-to-one correspondence of images in the training dataset.
We evaluate the generated outputs by discriminating them using the discriminator neural network, assuming that these outputs are similar to the training samples.
The discriminator model was trained using the same training dataset as that fed to the Boltzmann machine to learn its various features.
The discriminator then indicated the similarity between the generated and training samples.

The remainder of this paper is organised as follows. The next section describes the Boltzmann machine and sampling using a quantum annealer. 
Then, we explain the training and generation method of the Boltzmann machine used in the experiments and the discriminator to assess the generated data. 
Subsequently, we report the evaluation results of the images generated by the Boltzmann machine and compare different combinations of sampling methods in Boltzmann machine learning. 
In the last section, we summarise the study.

\section*{Problem Setting}
The Boltzmann machine is a neural network model with fully connected and undirected edges.
We set a binary variable that can be either $0$ or $1$ at each node. 
In Boltzmann machine learning, the binary variables are assumed to follow the Gibbs--Boltzmann distribution:
\begin{eqnarray}
\label{prob_boltz}
p({\bf x}|\boldsymbol{\theta}) &=& \frac{1}{Z(\theta)} \exp{(-\Phi({\bf x},\boldsymbol{\theta}))}, \\
\label{energy_boltz}
\Phi({\bf x},\boldsymbol{\theta}) &=& - \sum_{i}b_ix_i - \sum_{i,j}w_{ij}x_ix_j, \\
Z(\boldsymbol{\theta}) &=& \sum_{x}\exp(-\Phi({\bf x},\boldsymbol{\theta})) ,
\end{eqnarray}
where $x_i$ is the binary variable for node $i$ and $\Phi$ is the energy function of the Boltzmann machine indicating that the state of the node that reduces this energy is likely to appear. 
In the energy function, $b_i$ is the bias of node $i$ and $w_{ij}$ is the weight between nodes $i$ and $j$ on each edge. 
These are summarised as parameters and are denoted as $\boldsymbol{\theta}$. $Z(\boldsymbol{\theta})$ is a partition function used for normalisation.
In Boltzmann machine learning, the output data are assumed to follow the Gibbs--Boltzmann distribution defined above.
During learning, the maximum likelihood estimation for probability distribution ({\ref{prob_boltz}}) is performed:
\begin{eqnarray}
\ln{L(\boldsymbol{\theta})} &=& \sum_{k=1}^N \ln{p(\bf{x}^{(k)}|\boldsymbol{\theta})}, 
\end{eqnarray}
where 
\begin{eqnarray}
L(\boldsymbol{\theta}) &=& \prod_{k=1}^N p(\bf{x}^{(k)}|\boldsymbol{\theta})
\end{eqnarray}
and $N$ is the number of samples. 
We use the log-likelihood function instead of the mature quantity for simplicity.
To find maximiser $\boldsymbol{\theta}^*$ of the log-likelihood function, we take its derivative with respect to $\boldsymbol{\theta}$ to obtain 
\begin{eqnarray}
\frac{1}{N} \frac{\partial \ln L(\theta)}{\partial b_i} &=& \langle x_i \rangle_{\rm{data}} - \langle x_i \rangle_{\rm{model}}, \\
\frac{1}{N} \frac{\partial \ln L(\theta)}{\partial w_{ij}} &=& \langle x_{ij} \rangle_{\rm{data}} -\langle x_{ij} \rangle_{\rm{model}},
\end{eqnarray}
where $\langle \ldots \rangle_{\rm{data}}$ denotes the empirical mean over the training data and $\langle \ldots \rangle_{\rm{model}}$ denotes the model expectation.
The bottleneck of Boltzmann machine learning is estimating the model expectation in the equations above, as it requires $2^N$ calculations in principle. 
Therefore, the estimation should be approximated or a sampling method following the Gibbs--Boltzmann distribution should be adopted. 
For sampling, we prepared several synthetic data samples according to the probability distribution of the model and approximated the expected value using their empirical mean. 
Sampling methods include Gibbs sampling using the Markov chain Monte Carlo (MCMC) method.
Instead of directly manipulating Gibbs sampling, we used a quantum annealer to efficiently obtain output samples following the Gibbs--Boltzmann distribution.
We also compared the performance of Boltzmann machine learning for two sampling methods.

\begin{figure}[H]
\begin{tabular}{ccccc}
\begin{minipage}[b]{0.2\textwidth}
   \includegraphics[scale=0.3]{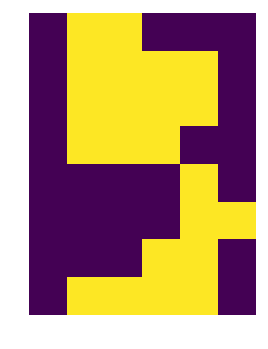}
   \centering
    \label{fig:n5}
  \end{minipage}
\begin{minipage}[b]{0.2\textwidth}
\includegraphics[scale=0.3]{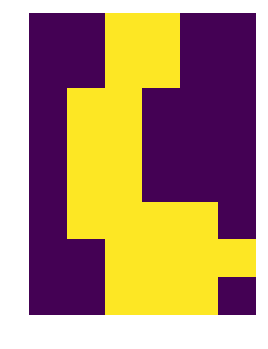}
\centering
\label{fig:n6}
\end{minipage}

\begin{minipage}[b]{0.2\textwidth}
   \includegraphics[scale=0.3]{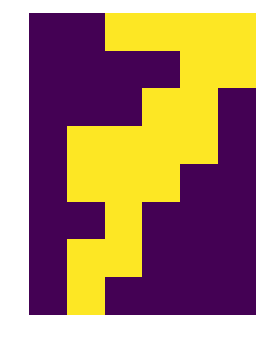}
   \centering
    \label{fig:n7}
  \end{minipage}
  
\begin{minipage}[b]{0.2\textwidth}
   \includegraphics[scale=0.3]{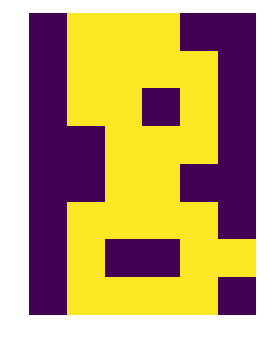}
   \centering
    \label{fig:n8}
  \end{minipage}

\begin{minipage}[b]{0.2\textwidth}
   \includegraphics[scale=0.3]{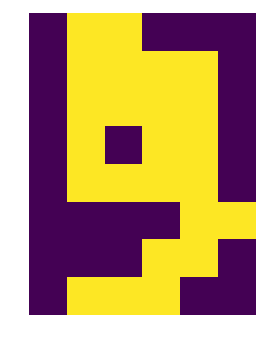}
   \centering
    \label{fig:n9}
  \end{minipage}
\end{tabular}
\caption{Samples of binarised handwritten digits from MNIST dataset.}
\label{digits}
\end{figure}

We analysed the quality of generation by the Boltzmann machine trained using the outputs from a quantum annealer.
To control data generation, we separated the nodes of the Boltzmann machine into sectors $A$ and $B$.
One sector of nodes ($i \in A$) represents the data, and the other sector ($i \in B$) selects the kind of data.
We denote the binary vectors in sector $A$ as ${\bf x}_A$ and those in sector $B$ as ${\bf x}_B$.
We used the MNIST dataset, which consists of various images of handwritten digits from zero to nine, as the dataset for assessment.
Thus, sector $B$ for selecting the digit label requires at most 10 nodes.
Sector $A$ representing the generated images has the remaining nodes of the Boltzmann machine.
Although machine learning studies are generally performed for higher-dimensional data even for proofs of concept, we employed the MNIST dataset due to the limited capacity of the quantum annealer.
For the experiments, we used the D-Wave 2000Q quantum computer, whose size is restricted to $64$ binary variables on the fully connected graph by embedding on a chimaera graph.
Consequently, the number of nodes in the sector for representing the image in Boltzmann machine learning was seriously restricted using this hardware.
We also restricted the number of digits to five, and resized the images to $8 \times 6 = 48$ pixels from the scaled-down MNIST dataset. 
In addition, we binarised the original images as shown in Fig. \ref{digits} to be suitable for the quantum annealer.

To control the generated images, we fixed one of the nodes in sector $B$ during training.
For instance, let us consider the generation of five images for digits from $5$ to $9$.
We represented each image ${\bf x}$ with label $k$ from $5$ to $9$ into image data ${\bf x}_{B} = {\bf x}$ using one-hot encoding as ${\bf x}_A = (1,0,0,0,0)$ for $k=5$, ${\bf x}_A = (0,1,0,0,0)$ for $k=1$, and so on.
The Boltzmann machine was trained using the labelled encoded images and provided the controlled images depending on sector $B$.
In the implementation, we applied a strong magnetic field to the node corresponding to the desired label.
Then, the generated output from the Boltzmann machine was conditioned on sector $B$.
In the conventional MCMC approach, we can directly control each bit in sector $B$.
In contrast, the quantum annealer cannot keep each bit.
Thus, we applied a strong magnetic field to generate the data.
Below, we detail the experiments for assessing the performance of Boltzmann machine learning by using a quantum annealer.
Note that the quantum annealer sometimes fails to correctly generate outputs by indirect control of the bit in sector $B$.

\section*{Experiments}
The currently available quantum annealer performs QA mainly to solve optimisation problems.
However, it can also generate different binary variable configurations.
The distribution of the configurations approximately follows a Gibbs--Boltzmann distribution with the form of the energy function in Eq.~(\ref{energy_boltz}). 
Therefore, we replaced the estimation of the model expectation with the computation of the empirical mean of the QA outputs by setting the same energy function during Boltzmann machine learning.
The outputs are classical but follow the Gibbs--Boltzmann distribution with a finite transverse magnetic field controlling the strength of the quantum fluctuation \cite{Amin2015}.
For optimisation, the remaining quantum fluctuation may lead to obstacles in the solutions.
However, for machine learning, it would be helpful to find the parameters providing high performance, as reported in previous studies.
Indeed, the possibility of improving the generalisation performance by finite-strength quantum fluctuations during neural network training has been explored \cite{Ohzeki2019,Arai2021}.

A previous study used the KL divergence to assess the performance of a Boltzmann machine \cite{Amin2018}.
As a result, a lower KL divergence was attained by using the Boltzmann machine and sampling from the quantum annealer, suggesting the superiority of the quantum version of the Boltzmann machine.
In this study, we focused on the practical performance of a quantum annealer by directly assessing the quality of generated data from the Boltzmann machine. 
Besides investigating the data quality, we compared different combinations of sampling methods for Boltzmann machine learning.
Specifically, we evaluated Gibbs sampling and direct sampling in the quantum annealer during both training and generation.
During training, sampling was used to compute the expectations for assessing the derivatives with respect to the parameters.
Finite-strength quantum fluctuations affect the precision of the computation of the derivatives.
In other words, quantum fluctuations perform regularisation,
as investigated in a previous study.
During generation, sampling is used again to generate a new sample.
Then, the finite-strength quantum fluctuation was assumed not to be directly related to the quality of the generated data but rather to cause degradation.
However, sampling in a quantum annealer provides independent output samples because the annealer quickly repeats the generation of the binary configurations following the Gibbs--Boltzmann distribution by leveraging quantum superposition. In contrast, classical sampling tends to maintain correlation between samples if both the generation of the distribution function is not restarted using different initial conditions and the sampling period is short.
Thus, Boltzmann machine learning using quantum annealers is expected to be superior to that using classical computers during generation.

\begin{figure}[H]
\centering
\includegraphics[scale = 0.2]{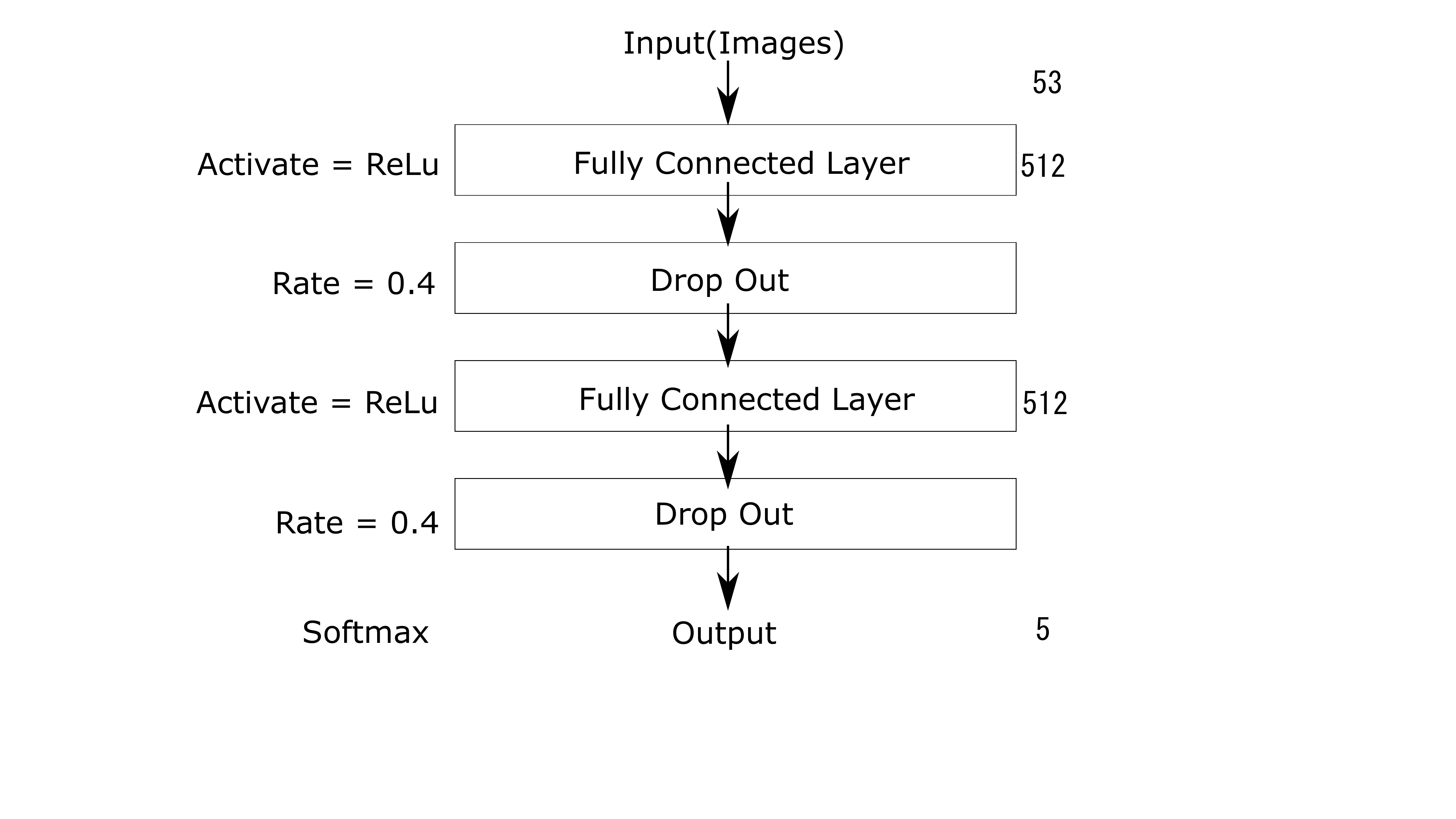}
\caption{Discriminator architecture (ReLU, rectified linear unit).}
\label{NN}
\end{figure}

We evaluated the performance of Boltzmann machine learning using a quantum annealer by preparing a discriminator neural network to measure the quality of the generated data.
The discriminator was trained using the MNIST dataset, as for Boltzmann machine learning.
The input layer of the discriminator had $8 \times 6 + 5 = 53$ nodes receiving the output from the Boltzmann machine representing the resized images and the bits in sector $B$ for selecting the label. The output layer had $5$ nodes to express the image label using one-hot encoding.
The discriminator architecture is shown in Fig. \ref{NN}. 
In addition, we set the cross-entropy as the loss function and used Adam optimisation to train the discriminator \cite{Kingma2015} and $D=896$ images representing digits 5--9. 
We used all these images for training the discriminator and the Boltzmann machine.
Thus, both networks suitably characterised the MNIST dataset.
However, the discriminator was not trained using the images generated by the Boltzmann machine.
We defined the quality of the generated images as the agreement rate between the obtained labels from discrimination: 
\begin{eqnarray}
R_a = \frac{D_m}{D_t},
\end{eqnarray}
where $D_m$ is the number of matching labels between generation and discrimination and $D_t$ is the total number of generated images.

We used two sampling methods, namely, Gibbs sampling and direct sampling in the quantum annealer (D-Wave 2000Q) for training. 
Gibbs sampling was performed with burn-in time $T_{\rm in}$ and sampling interval $\Delta t$ of $(T_{\rm in},\Delta t) = (200, 10)$ and $(1000, 50)$.
The number of samples was set to $200$.
On the other hand, direct sampling in the quantum annealer does not have burn-in time and sampling interval but uses annealing time, this we set to $\delta t = 5$ $\mu$s per sampling. The number of samples was $200$, as for Gibbs sampling. 
During training, while computing the expectation by the empirical mean trough sampling, we updated the parameters using minibatch learning and the momentum method with $L_2$ norm regularisation following the approach in \cite{Benedetti_2017}. 
The parameters used during training of the Boltzmann machine are listed in Table \ref{parameter}.

\begin{table}[H]
\centering
\caption{Parameters of Boltzmann machine.}
\label{parameter}
\begin{tabular}{c|c}
Parameter & Value \\ \hline \hline
Number of samples (images) & $896$ \\
Batch size & $256$ \\
Number of epochs & $4000$ \\
Number of samples per iteration & $200$ \\
Annealing time & $5$ $\mu s$ \\
Learning rate & $0.025$ \\
L2 norm regularisation & $0.00001$ \\
Momentum & $0.5$ \\
Quantum computer & D-Wave 2000Q\_5
\end{tabular}
\end{table}

\begin{figure}[H]
\centering
\includegraphics[scale = 0.2]{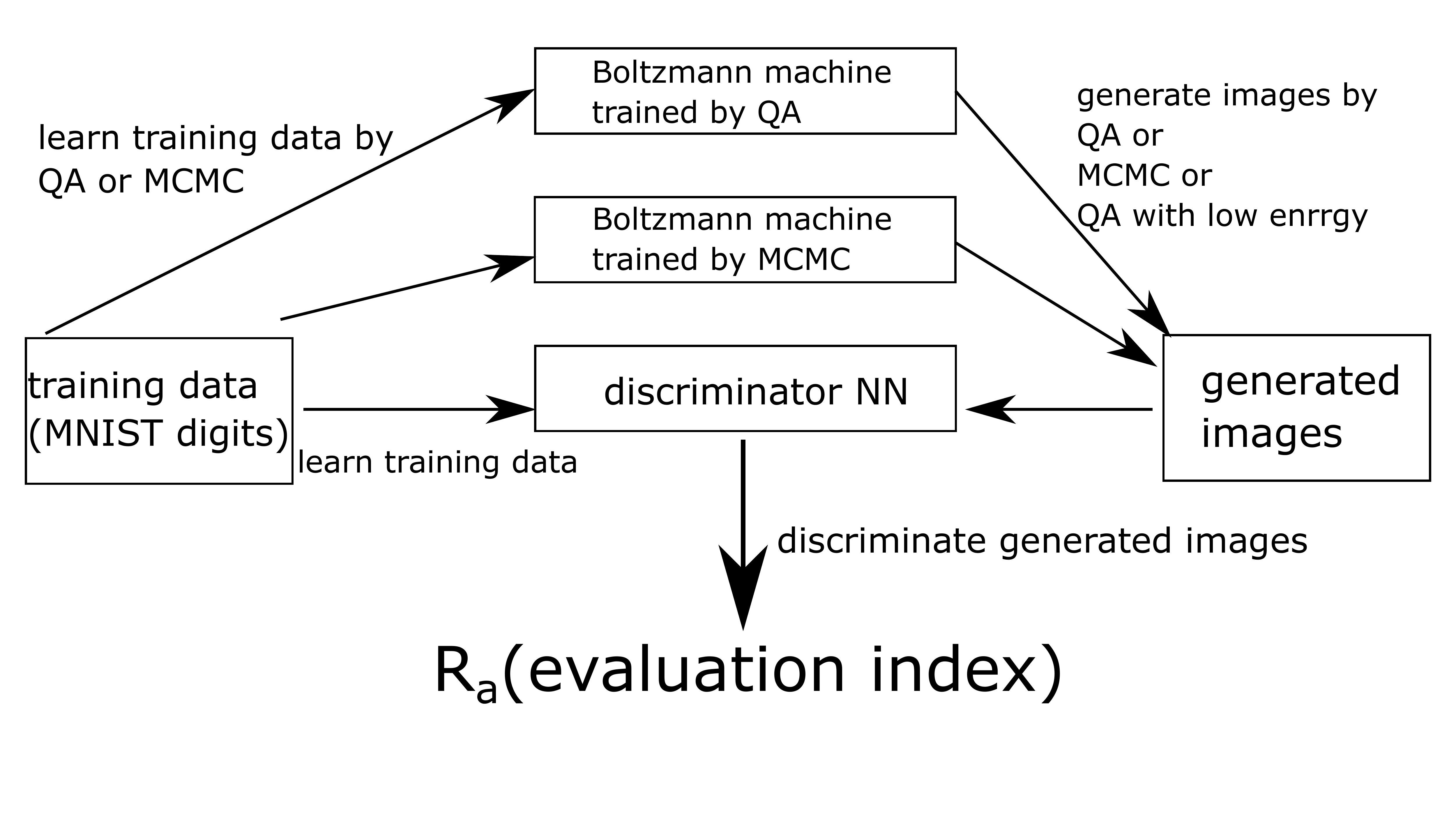}
\caption{Summary of experiments performed in this study.}
\label{experiments}
\end{figure}

After learning the Boltzmann machine, we used it to synthesise images.
For image generation, we also used Gibbs sampling and direct sampling in the quantum annealer. 
We sampled $100$ images per digit (i.e., $500$ samples for the five digits) by controlling sector $B$.
In addition to the two sampling methods, we performed low-energy sampling using the quantum annealer. 
Specifically, we first obtained $1000$ samples from the quantum annealer and considered only the $100$ images with the lowest energy values as the sampled images.
We changed combination of sampling methods for training and generation. 
For example, we considered training with Gibbs sampling and generation with direct sampling. The sampling combinations are listed in Table \ref{combination}.
\begin{table}[H]
\caption{Combinations of sampling methods for training and generation.
Gibbs sampling is denoted by MCMC, and lowQA denotes QA with the lowest energy states.}
\label{combination}
\centering
\begin{tabular}{ll}
Training & Generation \\ \hline \hline
QA & QA \\ \hline
QA & lowQA \\ \hline
QA & MCMC \\ \hline
MCMC & MCMC \\ \hline
MCMC & QA \\ \hline
MCMC & lowQA \\ \hline
\end{tabular}
\end{table}
All the experiments performed in this study are summarised in Fig. \ref{experiments}.

\section*{Results}
\begin{figure}[H]
\begin{tabular}{cc}

\begin{minipage}[b]{0.5\textwidth}
   \includegraphics[scale=0.5]{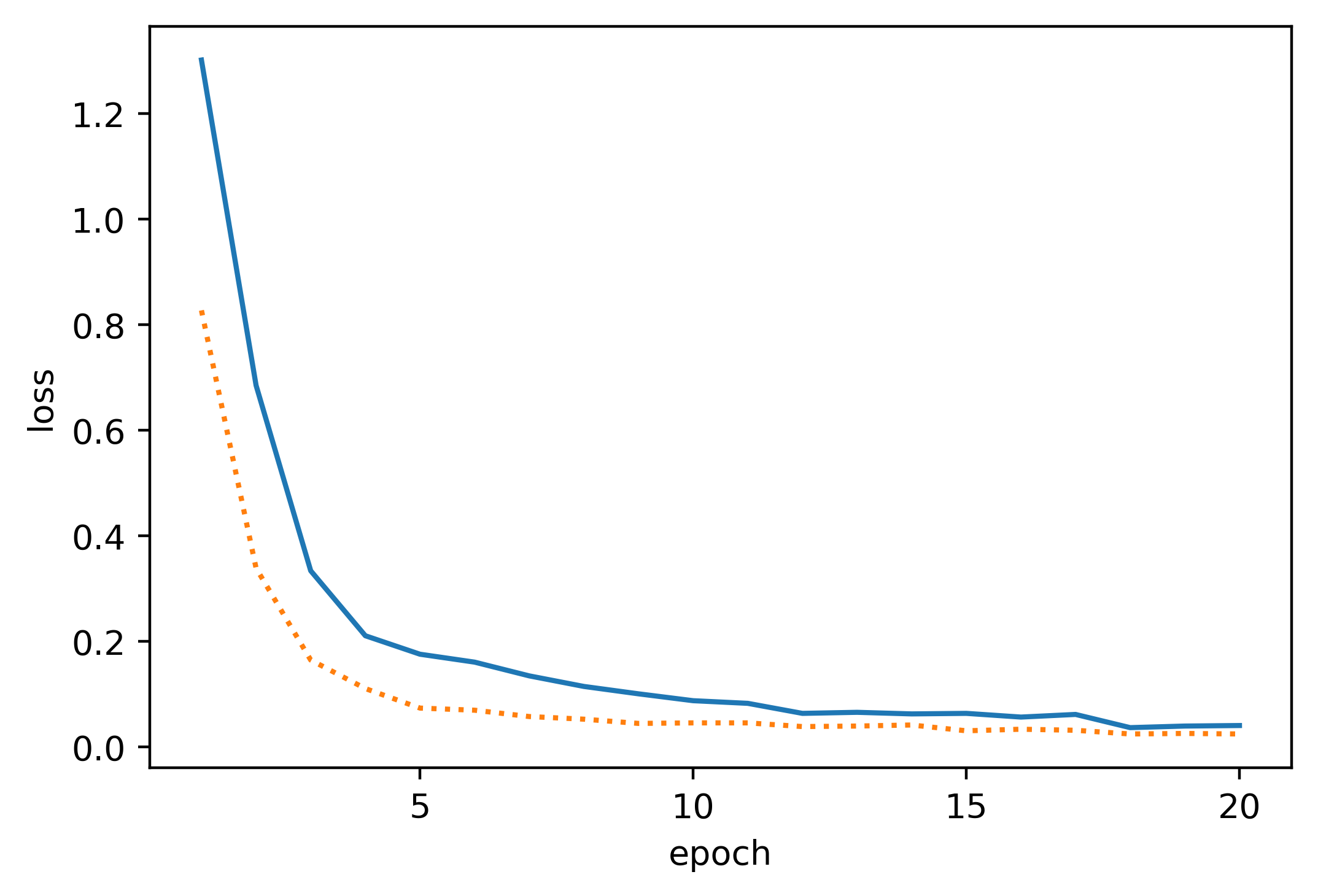}
   \centering
    \subcaption{Training loss}
    \label{fig:loss_discriminator}
  \end{minipage}
  
  \begin{minipage}[b]{0.5\textwidth}
   \includegraphics[scale=0.5]{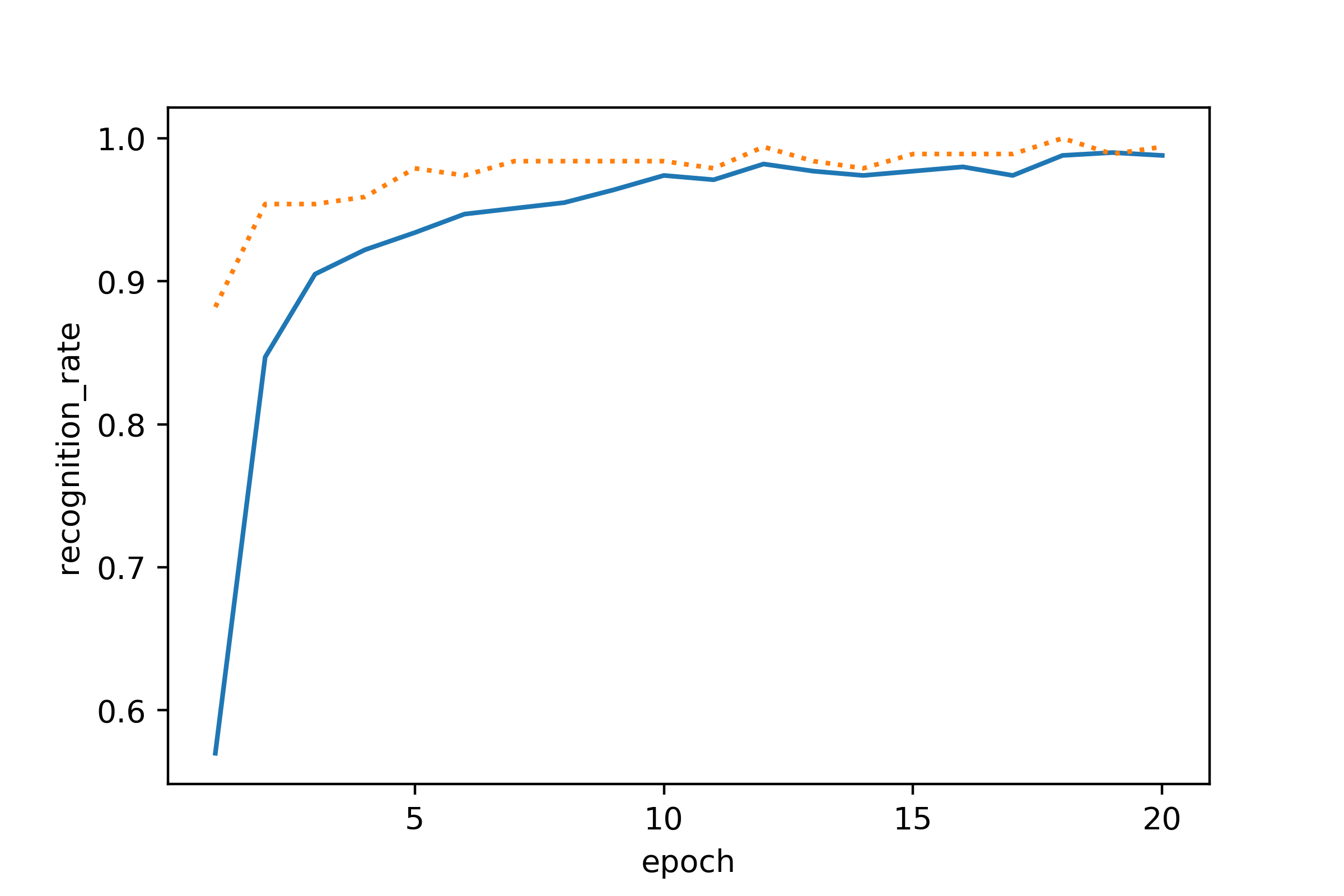}
   \centering
    \subcaption{Training recognition rate}
    \label{fig:recognition rate_discriminator}
  \end{minipage}
  
\end{tabular}
\caption{Results of discriminator training. 
The solid and dotted lines correspond to the results of training and validation, respectively.}
\label{discriminator}
\end{figure}

\begin{figure}[H]
 \begin{tabular}{cc}
\begin{minipage}[b]{0.5\textwidth}
   \includegraphics[scale=0.5]{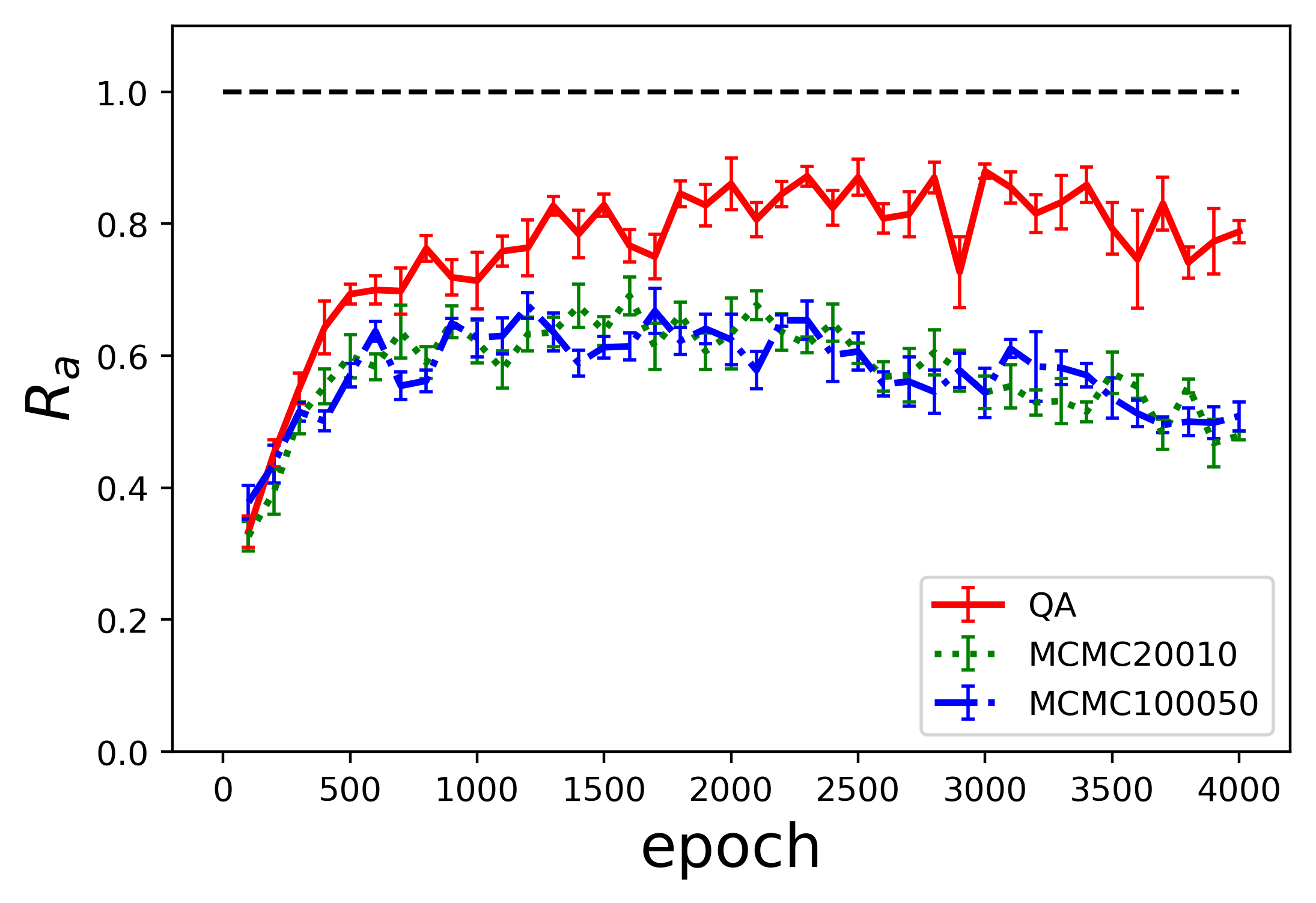}
   \centering
    \subcaption{QA generation}
    \label{fig:agreement_rate_QA}
  \end{minipage}
  \begin{minipage}[b]{0.5\textwidth}
   \includegraphics[scale=0.5]{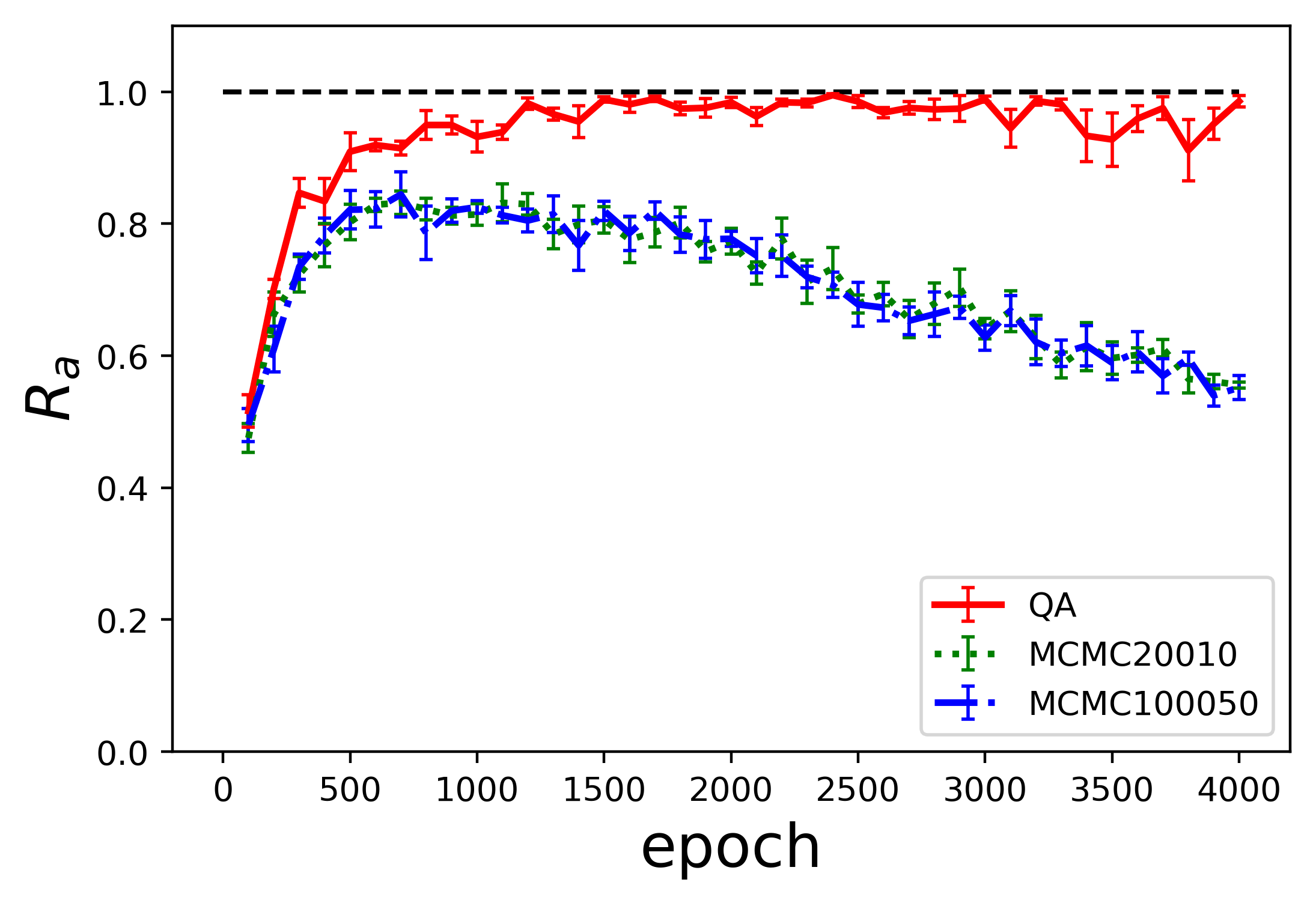}
   \centering
    \subcaption{Low-energy QA generation}
    \label{fig:agreement_rate_QAtop}
  \end{minipage}\\
\begin{minipage}[b]{0.5\textwidth}
    \includegraphics[scale=0.5]{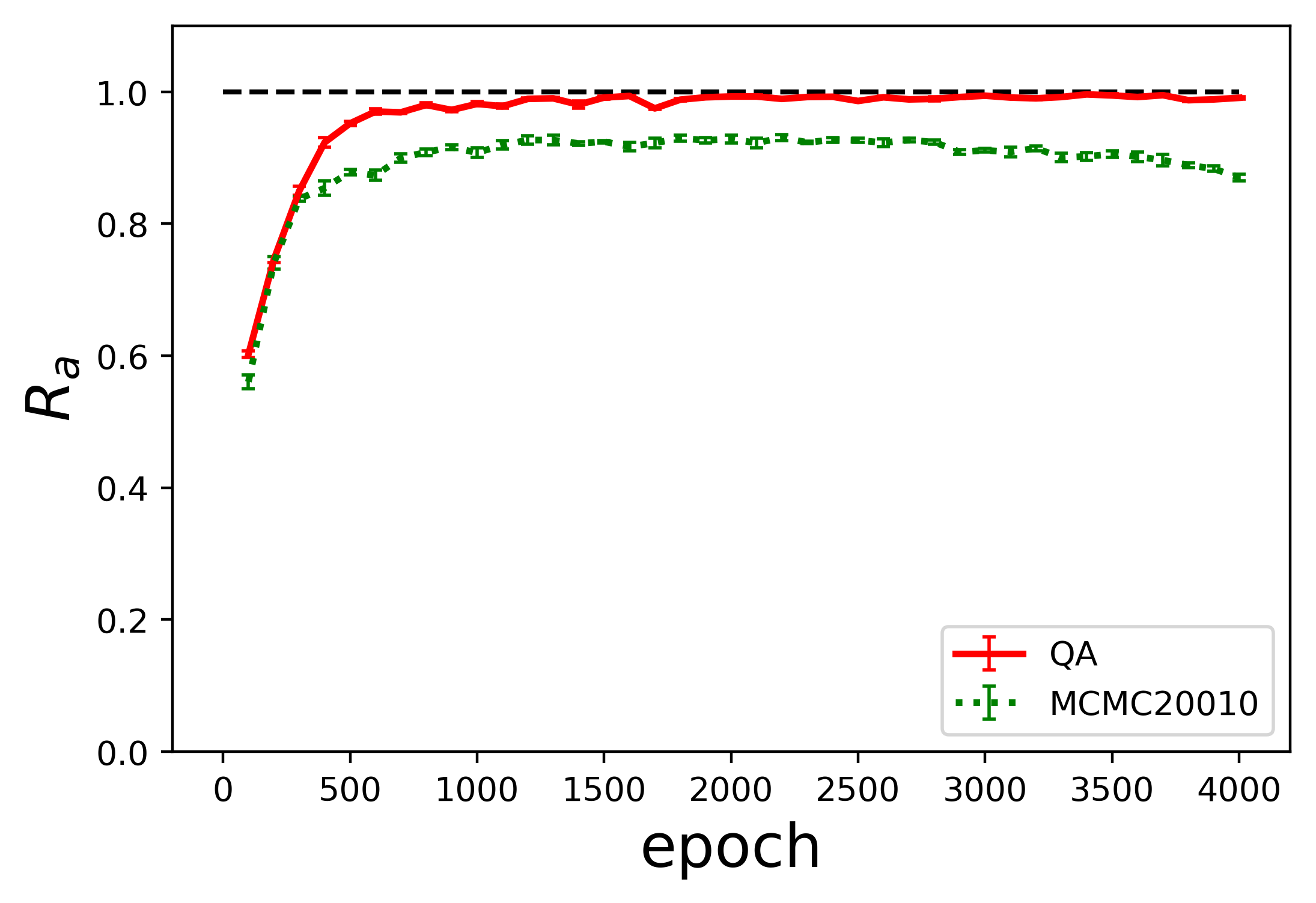}
    \centering
     \subcaption{MCMC($200,10$) generation}
     \label{fig:agreement_rate_MCMC20010}
  \end{minipage}
  \begin{minipage}[b]{0.5\textwidth}
    \includegraphics[scale=0.5]{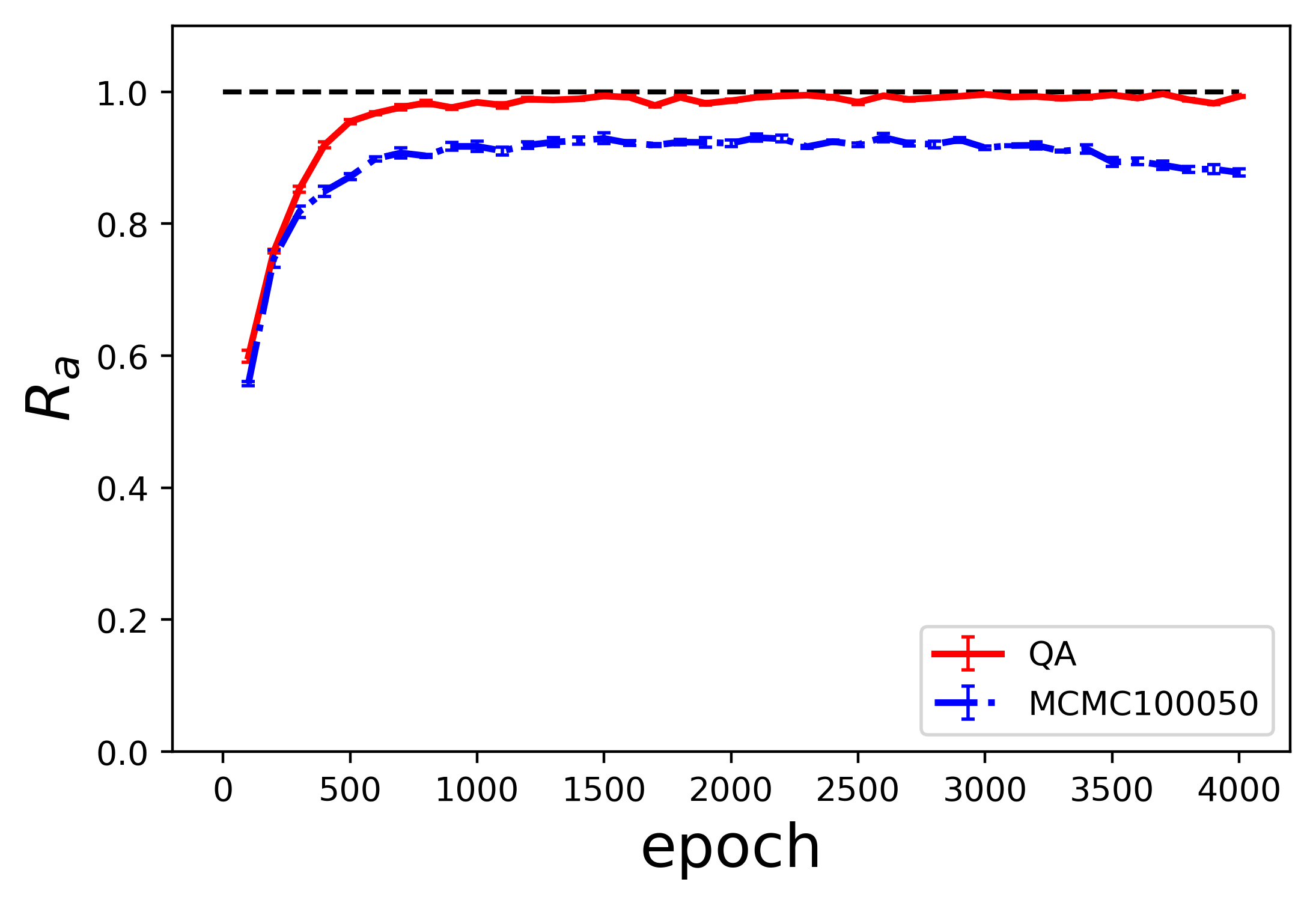}
    \centering
     \subcaption{MCMC($1000,50$) generation}
     \label{fig:agreement_rate_MCMC100050}
  \end{minipage}
 \end{tabular}
 \caption{Agreement rate $R_a$ over $4000$ epochs for different sampling methods for image generation. Training used direct sampling with QA (solid line), MCMC $(T_{\rm in},\Delta t)=(200,10)$ (dotted line), and MCMC $(T_{\rm in},\Delta t)=(1000,50)$ (dashed-dotted line).}
 \label{agreement rate}
\end{figure}

The training results of the discriminator is shown in Fig. \ref{discriminator}.
The recognition rate of the discriminator exceeded 99\%.
We used the discriminator to assess the performance of the Boltzmann machine
in terms of agreement rate $R_{a}$ according to the number of epochs, obtaining the results shown in Fig. \ref{agreement rate}.
For a given sampling method for generation, training using Gibbs sampling provided a $R_a$ approximately $10$\% lower than training using direct sampling with QA. 
When training using Gibbs sampling, $R_a$ tended to be lower owing to overfitting with increasing number of epochs. This phenomenon did not occur for direct sampling with QA. 
Thus, training using direct sampling with QA may prevent overfitting in training of the Boltzmann machine. 
This observation is consistent with previous studies that considered KL divergence \cite{Amin2018}. We confirmed the observation considering the agreement ratio.
The agreement ratio reflects the quality of the generated images by comparing the generated and original images.
On the other hand, the KL divergence measures the similarity between probability distributions.
Therefore, the agreement ratio is homogeneous between different generative models and measures the performance of data generation.

\begin{figure}[H]
\begin{tabular}{cc}

\begin{minipage}[b]{0.5\textwidth}
   \includegraphics[scale=0.3]{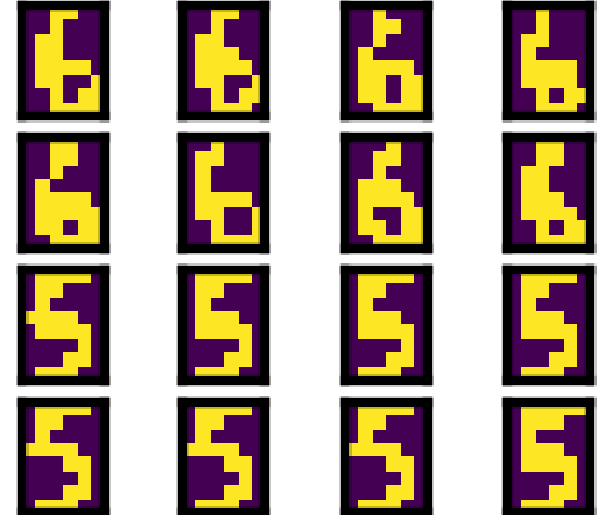}
   \centering
    \subcaption{Trained using direct sampling with QA}
    \label{fig:img trained by QA generated by QA with low energy}
  \end{minipage}
  
  \begin{minipage}[b]{0.5\textwidth}
   \includegraphics[scale=0.3]{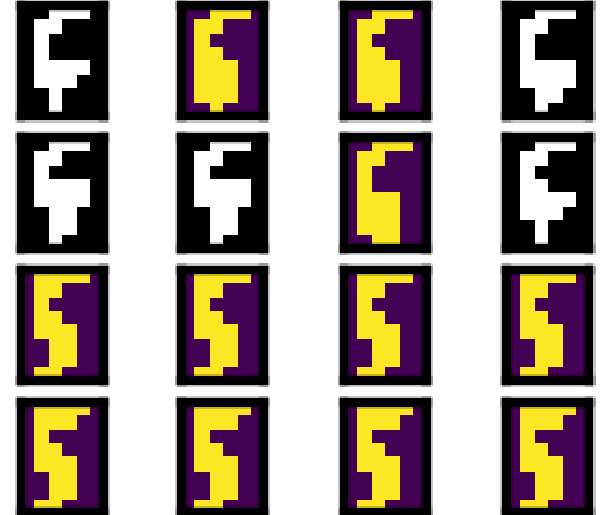}
   \centering
    \subcaption{Trained using MCMC($200,10$)}
    \label{fig:img trained by MCMC20010 generated by QA with low energy}
  \end{minipage}

\end{tabular}
\caption{Images generated using QA with low energy at epoch $4000$.}
\label{generated images QA}
\end{figure}

\begin{figure}[H]
\begin{tabular}{cc}
\begin{minipage}[b]{0.5\textwidth}
   \includegraphics[scale=0.3]{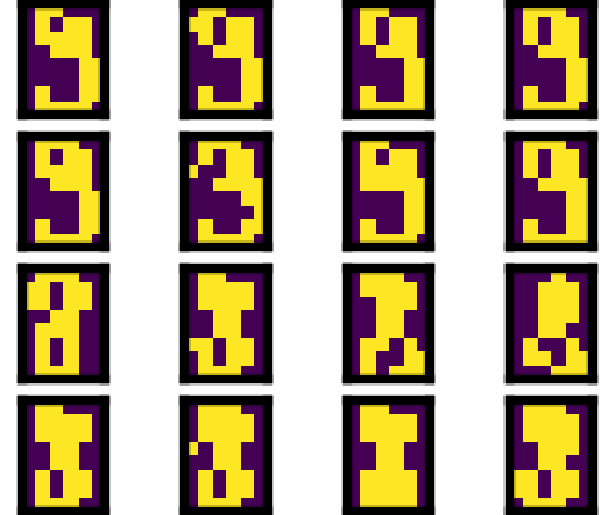}
   \centering
    \subcaption{Trained using direct sampling with QA}
    \label{fig:img trained by QA generated by MCMC20010}
  \end{minipage}
  
  \begin{minipage}[b]{0.5\textwidth}
   \includegraphics[scale=0.3]{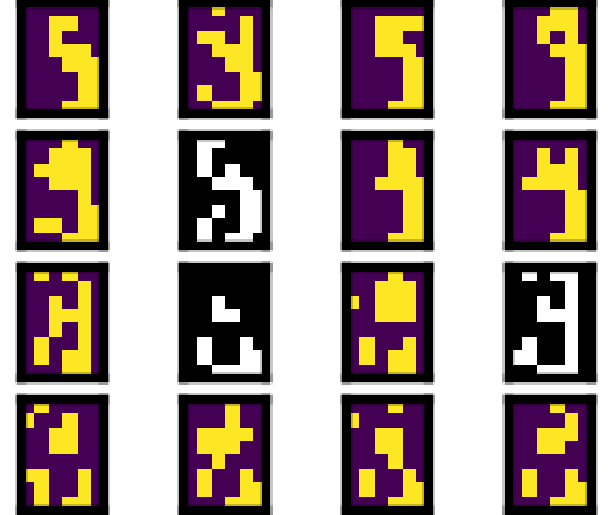}
   \centering
    \subcaption{Trained using MCMC($200,10$)}
    \label{fig:img trained by MCMC20010 generated by MCMC20010}
  \end{minipage}

\end{tabular}
\caption{Images generated using MCMC($200,10$) at epoch $4000$.}
\label{generated images MCMC}
\end{figure}

By considering the two characteristics of the agreement ratio, we compared all the experimental results.
Comparing the different sampling methods for training, we found that a higher agreement ratio is given by direct sampling with QA for all the cases regardless of the generation method.
Thus, regularisation is ensured during training of Boltzmann machine learning by direct sampling with QA.
Figs. \ref{generated images QA} and \ref{generated images MCMC} show several images to illustrate the differences between the training methods.
The images correctly discriminated are coloured, and those incorrectly classified are shown in black and white.

Comparing the different sampling methods during generation, we found that a higher agreement ratio is achieved by MCMC regardless of the burn-in time and sampling interval.
Thus, the quantum annealer provides lower generation quality than the classical approach.
This lower quality is due to obstacles in sector $B$ for selecting the label because we cannot always fix the binary variable during QA.
Instead, we applied a strong magnetic field to fix the binary variables in sector $B$.
Therefore, several samples could not be correctly generated following sector $B$.
This demonstrates some practical difficulty for generating data while controlling outputs using the quantum annealer.

\begin{figure}[H]
\begin{tabular}{cc}

\begin{minipage}[b]{0.5\textwidth}
   \includegraphics[scale=0.6]{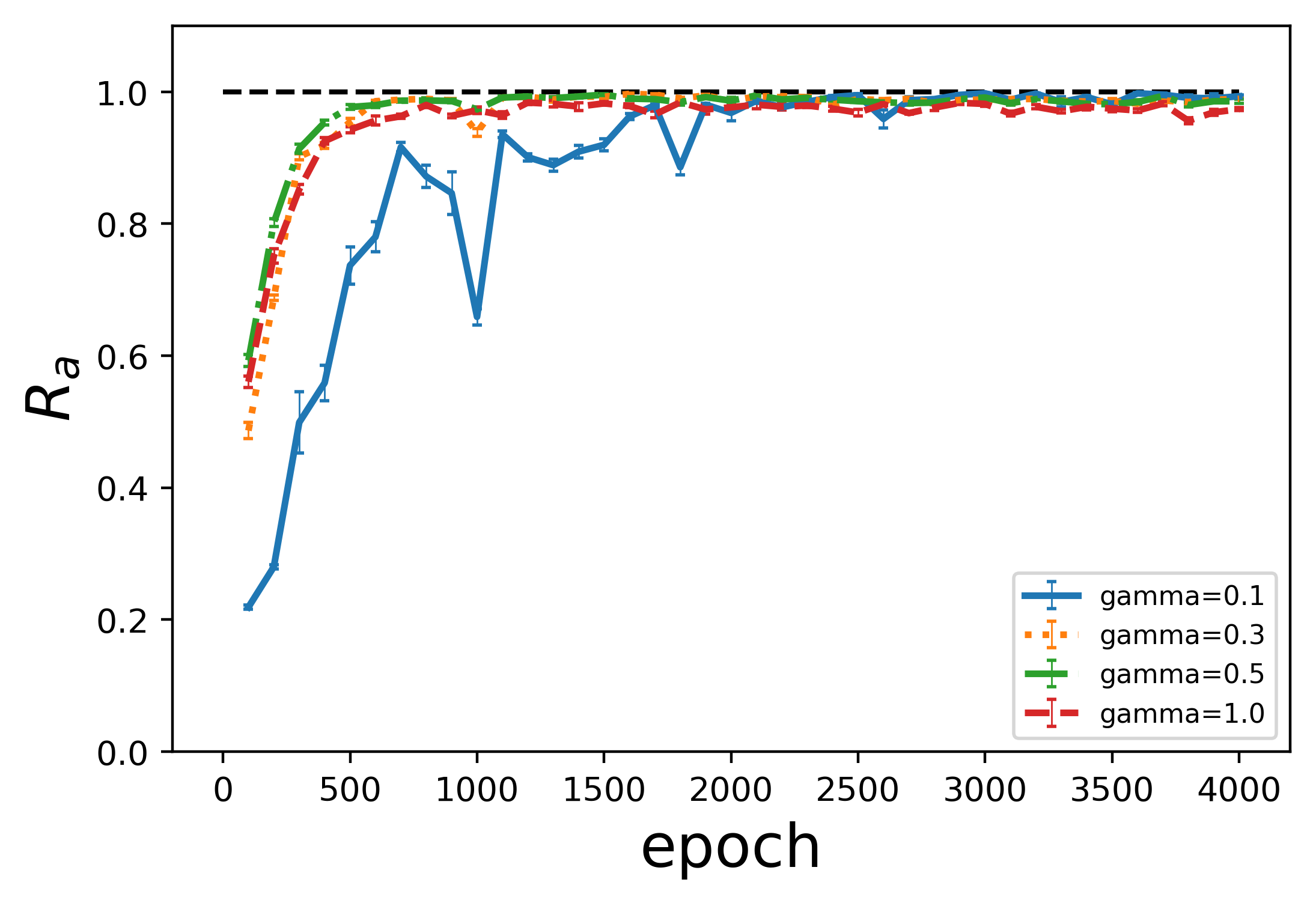}
   \centering
    \subcaption{Trained using direct sampling with QA}
    \label{fig:agreement rate trained by QA generated by QMC}
  \end{minipage}
  
  \begin{minipage}[b]{0.5\textwidth}
   \includegraphics[scale=0.6]{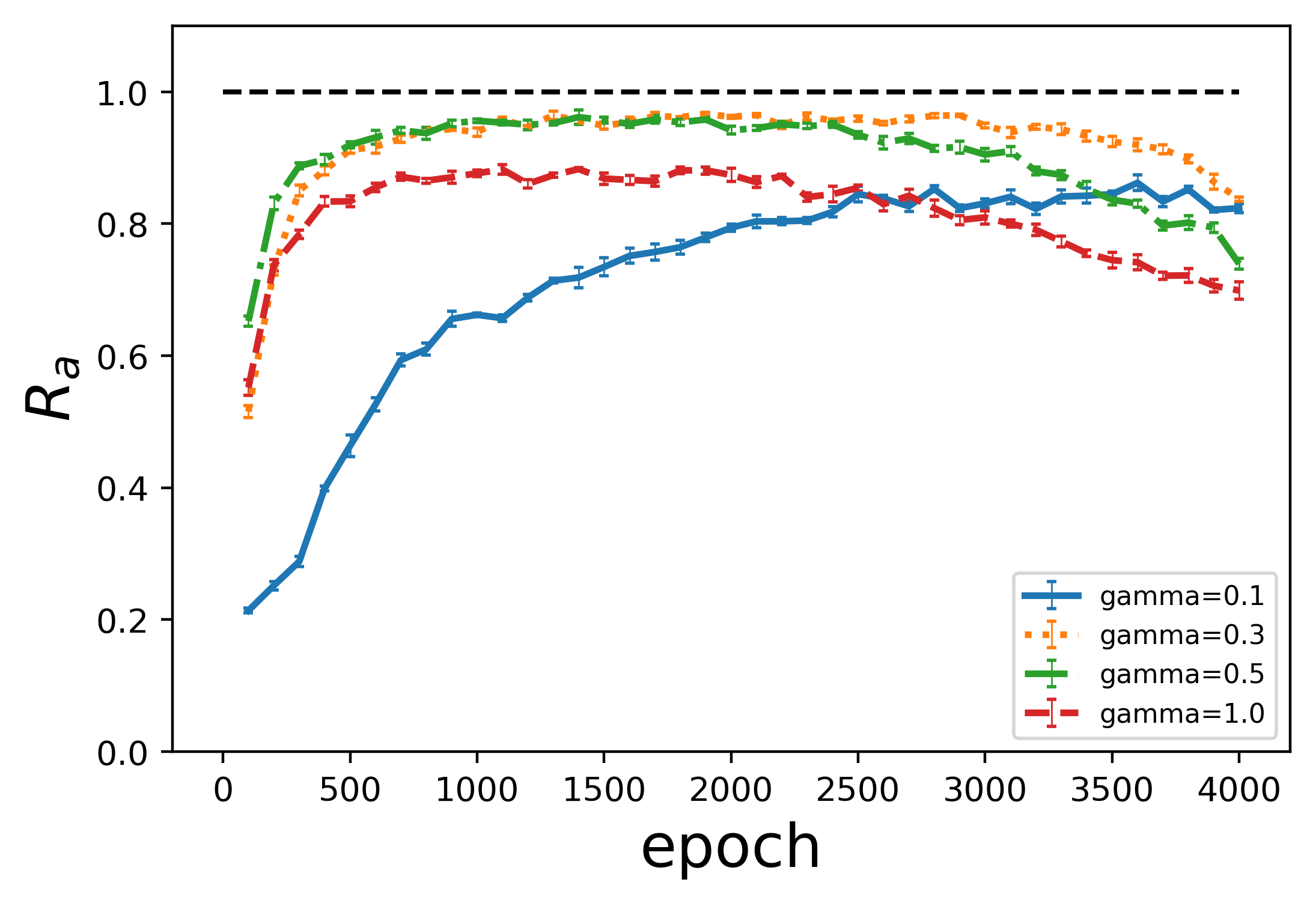}
   \centering
    \subcaption{Trained using MCMC($1000,50$)}
    \label{fig:agreement rate trained by MCMC100050 generated by QMC}
  \end{minipage}

\end{tabular}
\caption{$R_a$ over $4000$ epochs for images generated using quantum Monte Carlo method.}
\label{agereement rate QMC}
\end{figure}

From the experiments, it is clear that training using the quantum annealer improves the performance of the Boltzmann machine, being consistent with the findings from another study\cite{Amin2018}.
Regarding generation, QA retains a finite-strength transverse magnetic field at the end, influencing the experimental results. 
To investigate the effect of the residual transverse magnetic field, we applied the quantum Monte Carlo method with a finite-strength transverse magnetic field for data generation.
We used the same trained Boltzmann machines by QA and MCMC($1000,50$) as in the previous experiment. 
We set the number of Trotter slices to $32$ and the inverse temperature to $?$.
The transverse magnetic field was changed from $0.1$ to $1.0$ in increments of $0.1$ for image generation. $R_a$ was calculated according to the number of epochs during training, obtaining the results shown in Fig. \ref{agereement rate QMC}.

As shown in Fig. \ref{fig:agreement rate trained by QA generated by QMC}, the agreement rate is higher than when generating images using QA compared with the results shown in Fig. \ref{fig:img trained by QA generated by QA with low energy}.
Consequently, the performance reduces owing to the inherent effects of the quantum annealer and different control methods of sector $B$.
However, the strength of the transverse magnetic field does not improve because the performance saturates in this case.
During early training, a slight improvement was observed by inducing the finite-strength transverse magnetic field.
Thus, finite-strength quantum fluctuations may affect the quality of data generation, at least for the handwritten digits considered in this study.
In addition, considering generation using Gibbs sampling, we found substantial effects of the finite-strength transverse magnetic field.
The agreement rate considerably improved as the transverse magnetic field increased to $0.3$, but it reduced for a stronger field.
Therefore, a transverse magnetic field with appropriate strength may contribute to the performance improvement.
These results suggest that the improvement in $R_a$ by QA for generation using the Boltzmann machine is due to the slightly remanent transverse magnetic field.
Unfortunately, the performance of the quantum annealer hardware for generation is not better than that of Gibbs sampling for generation
due to the inherent effects affecting the hardware.
We did not analyse the degradation in performance of the Boltzmann machine for data generation.
Nevertheless, the last experiment demonstrated that the finite-strength quantum fluctuation in the quantum annealer implementation at the end of the protocol can improve data generation.
In the future, the improvement of the quantum annealer can lead to high-quality data generation, as suggested by the results of the last experiment.

\section*{Summary}
In this study, we performed various experiments to compare different sampling methods during training and generation based on Boltzmann machine learning.
Previous studies have mainly focused on the performance of Boltzmann machine learning by evaluating the KL divergence during training.
We analysed the effects of different sampling methods during data generation.
First, we tested sampling methods, including Gibbs sampling, and the repetition of using a quantum annealer and lower-energy samples from the quantum annealer during training.
The performance of the generated images was measured by a discriminator neural network that classified the images into several labels.
The highest performance was observed for the quantum annealer during training.
This is possibly due to the stochastic output provided by the quantum annealer.
The gradient computed during training was assessed by the empirical mean of the output provided by the quantum annealer.
Thus, the stochastic outputs result in fluctuating gradients. Consequently, the training escaped from several plateau and saddle points in the energy space of the log-likelihood function.
This finding is consistent with observations from previous studies.

In this study, we performed additional comparisons of the sampling methods during generation.
QA with lower-energy samples showed a relatively stable performance compared with direct sampling by the quantum annealer.
In addition, Gibbs sampling exhibited higher performance than the other sampling methods.
Thus, we can conclude that Gibbs sampling provides the highest performance for generation.
Thus, we applied the quantum Monte Carlo method to realise the ideal computation in the quantum annealer, which is usually polluted by environmental effects.
Sampling with the quantum Monte Carlo method showed higher performance than Gibbs sampling.
This means that the best generation quality is given by ideal QA implemented by the quantum Monte Carlo method.
Unfortunately, the available quantum annealer cannot realise ideal QA and thus shows poor performance during generation.

We varied the strength of the transverse magnetic field and investigated the performance of generation for different sampling methods.
We found that a finite-strength transverse magnetic field enhances the quality of the generated images.
This is a remarkable discovery in the context of quantum machine learning.
Previous studies ensured that finite-strength quantum fluctuations during training improved the generalisation performance, indicating the robustness of the trained model and ability to handle unknown data.
On the other hand, we showed that finite-strength quantum fluctuations increase the quality of the generated images.
We could not assert any mechanism that explains the higher performance of sampling with finite-strength quantum fluctuations.
However, we believe that our findings provide a new direction for studying quantum machine learning.
The latest version of the quantum annealer, the D-wave advantage, might show different (possibly higher) performance compared with that achieved in our study.
As shown in the results of the quantum Monte Carlo method, sampling affected by the remanent quantum fluctuation at the end of the QA protocol showed a slightly different performance during generation compared with Gibbs sampling.
After this study, we expect that future developments will emphasise practical aspects of quantum machine learning, such as generated images and their quality.

\bibliography{QABoltzmannmachine_ja}

\section*{Acknowledgements}
The authors would like to thank Manaka Okuyama and Masamichi J. Miyama for the fruitful discussions. 
This work was financially supported by JSPS KAKENHI Grant No. 19H01095 and No. 18H03303, partly supported by JST-CREST (No. JPMJCR1402), the Next Generation High-Performance Computing Infrastructures and Applications R\&D Program of MEXT and by MEXT-Quantum Leap Flagship Program Grant
Number JPMXS0120352009.

\section*{Author contributions statement}
T. S. performed the experiments and analysed the results; M.O. led the research project and established the experimental design and method for evaluating the results; all the authors contributed to writing and improving the manuscript.

\section*{Additional information}
{\bf Competing Interests}: The authors have no competing interests to declare.

\end{document}